A System of Acquisition and Analysis of Data for Extraction of Knowledge of the Ebit Platform

Um Sistema de Aquisição e Análise de Dados para Extração de Conhecimento da Plataforma Ebit

Marcelo Augusto Muniz Teixeira, ( Universidade Estadual do Maranhão, MA, Brasil) - augustomnzteixeira@gmail.com

Fábio Manoel França Lobato, 0000-0002-6282-0368, (Universidade Federal do Oeste do Pará, PA, Brasil) - fabio.lobatof@ufopa.edu.br

Beatriz Nery Rodrigues Chagas, (Universidade Estadual do Maranhão, MA, Brasil) - beatriznery.12@gmail.com

Antonio Fernando Lavareda Jacob Junior, 0000-0002-9415-7265, (Universidade Estadual do Maranhão, MA, Brasil) - antonio.jacob@gmail.com

Abstract: The internet development and the consequent change in communication forms have strengthened as online social networks, increasing the involvement of people with this media and making consumers of products and services, which are more informed and demanding for companies. This context has given rise to Social CRM, which can be put into practice by means of electronic eletronic word of mouth platforms, enable web sharing of comments and evaluations about companies, defining their reputation. However, most electronic word of mouth platforms do not provide information for extracting your information, making it difficult to analyze the data. To satisfy this gap, a system was developed to capture and automatically summarize the data of the companies registered in the eBit platform.

Resumo: O desenvolvimento da internet e a consequente mudança das formas de comunicação fortaleceram as redes sociais *online*, aumentando o envolvimento das pessoas com este meio de comunicação e tornando os consumidores de produtos e serviços, os quais estão mais informados e exigentes para com as empresas. Este contexto deu origem ao CRM Social, que pode ser colocado em prática por meio das plataformas de boca a boca eletrônico, possibilitam o compartilhamento na *web* de comentários e avaliações sobre as empresas, definindo a sua reputação. No entanto, a maioria das plataformas de boca a boca eletrônico não oferece nenhuma interface para extração das suas informações, dificultando a análise dos dados. Para preencher esta lacuna, foi desenvolvido um sistema para capturar e sumarizar de forma automática os dados das empresas cadastradas na plataforma eBit.

Palavras-chave: Redes Sociais, Boca-a-Boca Eletrônico, CRM Social

# 1 INTRODUÇÃO

Com o desenvolvimento da internet e, sobretudo com a democratização das formas de acesso à mesma e o aumento da largura de banda, os meios de comunicação mudaram completamente, potencializando as redes sociais (Rosa, 2010). Por conta disso, há um engajamento cada vez maior das pessoas com as redes sociais (Fan & Gordon, 2014), tornando os consumidores de produtos e serviços mais informados e exigentes. Este fato está proporcionado que as empresas precisem encontrar novas formas de responder às suas necessidades (Matias, 2012).

Este contexto, trouxe à tona um novo conceito que possibilita a colaboração entre cliente e empresa, oferecendo valor mutualmente benéfico, o Social CRM (Küpper et al., 2014), que designa medidas que utilizam as tecnologias de redes sociais no planejamento, implementação e controle de atividades de CRM - *Customer Relationship Management* ou Gerenciamento de Relacionamento com o Cliente (Ang, 2011).

Uma forma de comunicação bastante utilizada no que diz respeito ao CRM Social é o boca-a-boca eletrônico, também conhecido como eWoM (Eletronic World-of-Mouth), que permite aos internautas a criação e o compartilhamento de conteúdos on-line que representam avaliações positivas ou negativas sobre um produto, marca ou serviço (Arndt, 1967; Tudesco, 2014). Uma empresa avaliada negativamente corre sérios riscos de ter sua reputação manchada e consequentemente, seu negócio prejudicado (Grégoire & Fisher, 2006), pois, por meio da Internet, o consumidor é capaz de expor sua indignação a um grande número de pessoas (Hennig-Thurau et al., 2004). Este fato vem obrigando as empresas a melhorarem a qualidade dos serviços prestados, com o intuito de evitar este transtorno.

Para que as pessoas possam registrar e acessar informações referentes à qualidade das empresas, existem os sistemas de reputação. Este tipo de sistema tem como objetivo coletar e publicar informações a respeito do histórico do comportamento de vendedores dos mais variados tipos de mercado (Kennes & Schiff, 2007). Um exemplo de plataforma de boca-a-boca eletrônico bastante popular no Brasil é o eBit, cujo propósito é agregar reclamações, além de servir como um canal de comunicação entre consumidores e empresas. Porém, esta plataforma não disponibiliza nenhuma interface para capturar seus dados, dificultando o processo de análise e monitoramento.

É importante destacar que a análise destes dados pode ajudar a melhorar a qualidade da informação disponibilizada aos consumidores, tornando as interações cliente-empresa mais interessantes ao empregar lealdade, aumentando as chances de manter a fidelidade dos atuais e atrair novos clientes (Alt & Reinhold, 2012). Entretanto, há uma grande dificuldade em explorar plenamente os dados, devido ao seu grande volume, além de boa parte das informações estarem carregadas de desafios para uma máquina, entre eles, conteúdo com ambiguidade, ruídos e sentimentos expressos (Lima, 2017).

Dessa forma, o objetivo da presente pesquisa é implementar um sistema de aquisição e análise automática de dados capaz de ajudar os consumidores de produtos e serviços a obter conhecimentos acerca da reputação das empresas com base nas informações disponibilizadas pela plataforma de boca-a-boca eletrônico eBit, de modo a melhorar a experiência dos clientes. Para atingir tal objetivo criou-se dois algoritmos, sendo um para extrair os dados da plataforma eBit e outro para realizar a modelagem de tópicos sobre os dados extraídos. Além disso, desenvolveu-se uma interface *web*, para que as pessoas pudessem consultar informações sobre as empresas e salvar tais informações em um arquivo para posterior análise. Para isso, foi necessário programar esta interface para executar de forma automática os dois algoritmos propostos.

Este artigo encontra-se organizado como segue: a Seção 2 apresenta dois trabalhos relacionados a aquisição e análise automática de dados. Na Seção 3, são apresentados os métodos e o processo utilizado para o desenvolvimento do sistema proposto. Na Seção 4, os estudos de caso e seus resultados são descritos. Por fim, as Seções 5 e 6 apresentam as considerações finais e as recomendações de trabalhos futuros.

## 2  TRABALHOS RELACIONADOS

OLMEDILLA & MARTÍNEZ-TORRES (2016) desenvolveram um *web crawler* para coleta de dados do portal Ciao (www.ciao.com/), uma das maiores comunidades mundiais de boca-a-boca eletrônico. Este experimento é dividido em duas partes: o rastreamento dos dados gerados pelos usuários; e armazenamento dos dados coletados em um banco de dados. Para rastrear os dados do portal, foi utilizada a linguagem de programação Python combinada com o framework Scrapy. Na segunda etapa, foi criado um banco de dados MySQL para depositar os dados capturados.

ALMEIDA et al. (2017) implementaram um conjunto de métodos para analisar de forma automática os dados da plataforma ReclameAqui, um site brasileiro utilizado para registrar reclamações sobre as empresas, que além disso, possibilita a interação entre as empresas e os consumidores. Este projeto contou com duas fases principais. Em primeiro lugar, realizou-se a captura de dados de duas empresas cadastradas na plataforma. Por último foi feita a modelagem de tópicos dos dados coletados por meio da técnica LDA.

Os dois trabalhos apresentados têm como objetivo extrair conhecimento de numerosos conjuntos de dados não estruturados, presentes em plataformas de boca-a-boca eletrônico. Ambos tiveram êxito nas suas propostas.

O primeiro trabalho apresenta duas principais limitações. O método implementado não consegue extrair informações de páginas *web* feitas com Javascript. Além disso, como não é realizada nenhuma análise dos dados coletados, nenhum conhecimento é gerado.

O último trabalho preenche as lacunas deixadas pelo primeiro, tornando possível extrair conhecimento de um conjunto de dados, não importando a sua amplitude.

Ambas as abordagens foram de suma importância para o desenvolvimento do presente projeto.

## 3  METODOLOGIA

Para implementar o sistema foi preciso utilizar três métodos: *web crawler*, desenvolvimento web e modelagem de tópicos.

a.  Método de *web crawler*

Uma forma de realizar a aquisição dos dados da Plataforma eBit de forma automática é o emprego de w*eb crawler* ou rastreador *web*, um programa que navega na internet autonomamente, a partir de uma lista de semente de páginas *web* e visita recursivamente os documentos acessíveis a partir dessa lista. Seu principal objetivo é descobrir e recuperar conteúdo e conhecimento da *web* (Stevanovic et al., 2012).

Uma das tecnologias utilizadas para programar rastreadores *web* é o *Scrapy*, um *framework* de código aberto lançado em junho de 2008 e desenvolvido na linguagem de programação Python. Este framework possibilita a aquisição de informações de páginas web com alto desempenho, podendo ser utilizado para uma variedade de aplicações úteis, tais como mineração de dados, processamento de informações e arquivamento histórico.

Apesar de ter sido projetado inicialmente para raspagem de dados da internet, o Scrapy também pode ser utilizado para extração de dados usando APIs ou como um rastreador web de propósito geral (Alves, 2016; Zubeda et al., 2016).

Quanto à arquitetura e suas funcionalidades (Figura 1), o Scrapy apresenta seis componentes principais (Jailia et al., 2016):

- *Scrapy Engine*, responsável por controlar o fluxo de dados entre todos os componentes do sistema;
- *Scheduler*, encarregado de receber as requisições do *Scrapy Engine* e adicioná-las a uma fila para posterior atendimento, de acordo com as solicitações;
- *Downloader*, responsável por procurar páginas *web* e enviá-las para o *Scrapy Engine*, que, por sua vez, as repassará às *Spiders*;
- *Spiders*, também conhecidas como rastreadores *web*, classes implementadas para analisar as respostas e extrair itens delas ou URLs adicionais;
- *Item Pipeline*, responsável por processar os itens extraídos pelas *Spiders*;
- Os *middlewares* do *Downloader*, componentes específicos que processam solicitações enviadas do *Scrapy Engine* para o *Downloader* e as respostas que passam do *Dowloader* para o *Scrapy Engine*.

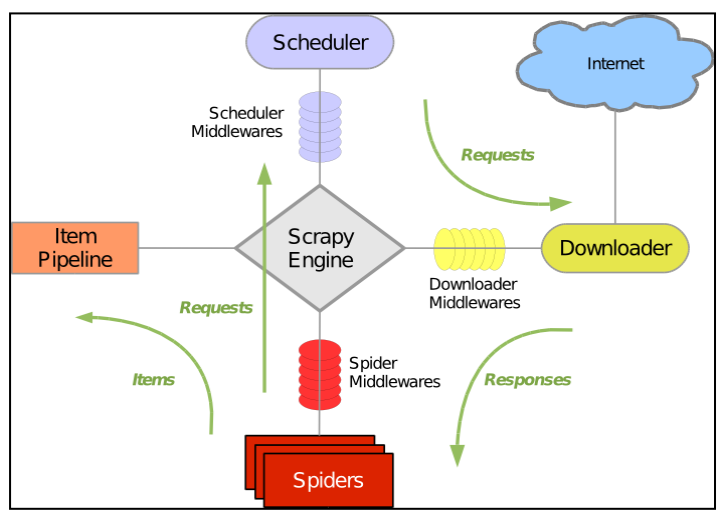

**Figura 1 - Arquitetura do Scrapy, Fonte: Wang & Guo, 2012**

Dentre os componentes citados, são as *spiders* que possuem relação direta com o programador, uma vez que a intervenção humana é necessária para definir quais páginas *web* serão acessadas, quais informações serão extraídas e como serão extraídas. Portanto, as *spiders* são classes que definem como um determinado *site* será rastreado, como o rastreamento será executado e a forma de extração dos dados estruturados das páginas.

b. Modelagem de tópicos

Após a aquisição dos dados, é realizada a análise dos mesmos para a extração de conhecimento por meio da técnica de modelagem de tópicos, a fim de enriquecer a informação obtida. A ideia desta técnica consiste em explorar uma coleção de dados discretos de modo a encontrar uma descrição dos seus elementos. Um dos benefícios dessa técnica é possibilitar o processamento de uma grande quantidade de dados (Jordão, 2016), permitindo a organização e a sumarização de arquivos eletrônicos a uma escala impossível para o ser humano (Blei, 2012).

Esta etapa pode ser realizada através do uso de LDA (Latent Dirichlet Allocation), que é um modelo de geração probabilístico, o qual pode ser utilizado para modelar e recuperar estruturas subjacentes de tópicos de qualquer tipo de dado discreto, como o tipo textual (Al-Salemi et al., 2014). A técnica LDA é uma das mais utilizadas em modelagem de tópicos e tem como base a ideia de que um documento é um aglomerado de tópicos (FRIAS, 2016), ou seja, é capaz de reconhecer tópicos ocultos em uma coleção de documentos (Blei & Jordan, 2003), havendo em cada tópico um grupo de palavras com uma determinada probabilidade de pertencer a ele. Portanto, para cada documento, há uma quantidade Nd de palavras, onde cada palavra foi obtida a partir de um dos k tópicos existentes (Neves, 2016).

c. Desenvolvimento *web*

O desenvolvimento *web* consiste em uma prática de desenvolvimento que envolve a *web*. São atividades que abrangem desde a criação de simples páginas de texto até o desenvolvimento de aplicações complexas. Neste trabalho, foi utilizada a ferramenta Django, um framework de desenvolvimento rápido para a Web implementado na linguagem de programação Python (Brito, 2016). Este framwework foi criado no ano de 2003 como um sistema para administrar um *site* de notícias *online*. Em 2005, tornou-se um *software* livre e, desde então, é mantido por uma comunidade de desenvolvedores (Erlo, 2011).

Quanto ao seu funcionamento, o Django é inspirado na arquitetura MVC (*model-view-controller*), que permite aos desenvolvedores a criação de suas aplicações separando a visualização interna da informação da forma como ela será exibida ao usuário do sistema (Jailia et al., 2016), conforme Figura 2.

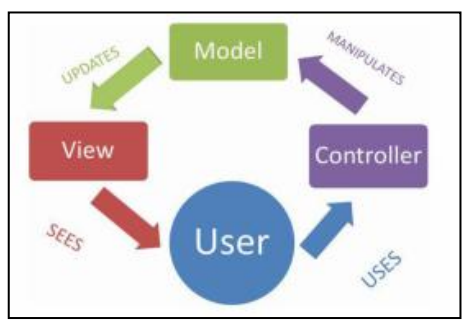

**Figura 2 - Arquitetura MVC, Fonte: Jailia et al., 2016**

Esta arquitetura é dividida em três camadas que interagem entre si:

- *Model*, responsável por gerenciar os dados do aplicativo e permitir a comunicação do banco de dados;
- *View*, encarregada de exibir ao usuário os dados da camada *Model* ocultando as informações desnecessárias;
- *Controller*, responsável por controlar as camadas *Model* e *View* e o fluxo de dados, atualizando a exibição quando os dados são alterados.

Apesar de ter como base a arquitetura MCV, o *framework* Django implementa um padrão de arquitetura similar, o MTV (*model-template-view*). Esta arquitetura possui quatro componentes (Cantú, 2013), conforme é exibido na Figura 3:

- URLs, responsáveis por disponibilizar os recursos e as rotas do sistema;
- *Model*, encarregado de organizar os dados do sistema, desconsiderando as informações desnecessárias;
- *Template*, cuja função é disponibilizar e exibir os dados ao usuário;
- *View*, que serve como um canal de comunicação entre *Model* e *Template*, recebendo requisições (*request*) e retornando respostas (*response*).

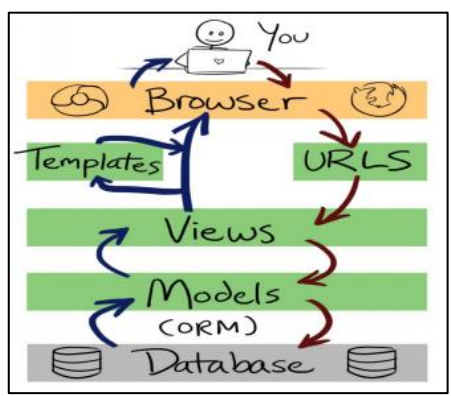

**Figura 3 - Arquitetura MTC, Fonte: Cantú, 2013**

d. Implementação do sistema

O processo de implementação do sistema foi dividido em três fases. Em primeiro lugar, programou-se o *web crawler* para capturar os dados da plataforma eBit. Em seguida, desenvolveu-se a interface *web* para interação com o usuário. Por último foi implementado o algoritmo LDA para sumarizar os dados extraídos. O funcionamento do sistema divide-se em oito etapas conforme é exibido na Figura 4, onde sua arquitetura é representada.

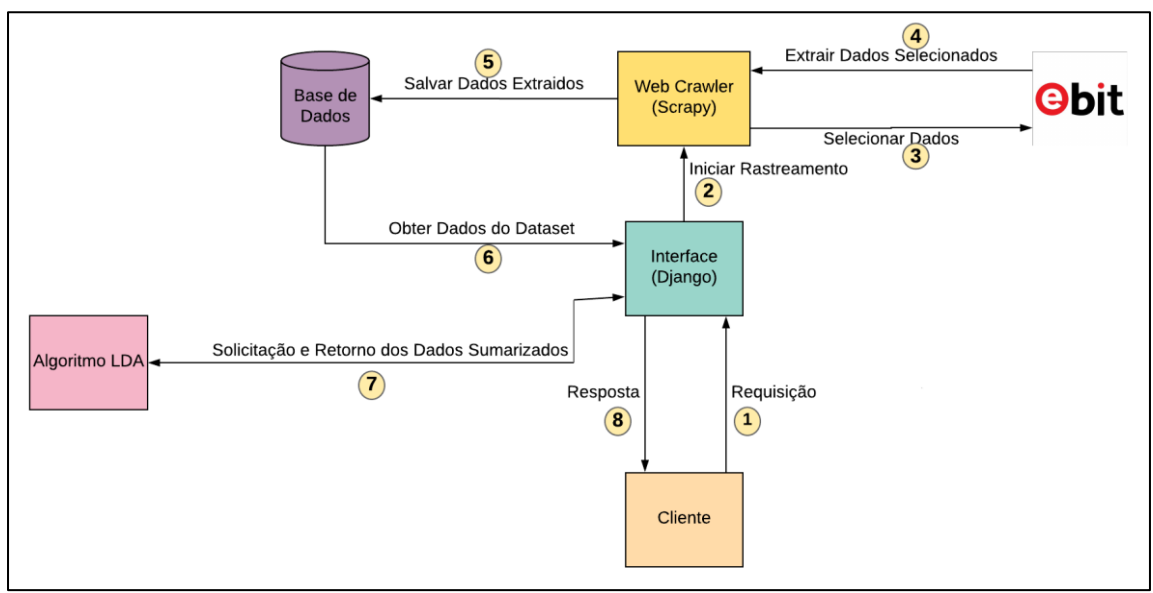

**Figura 4 - Arquitetura do sistema**

As etapas de execução do sistema são:

- Requisição (1): o Cliente envia uma solicitação para a Interface WEB (Django), ao inserir em um formulário o nome da empresa que ele deseja consultar e clicar no botão BUSCAR;
- Início do rastreamento (2): o Web Crawler (Scrapy) é acionado e começa a navegar na plataforma eBit, a fim de capturar os dados da empresa solicitada pelo cliente;
- Seleção dos dados (3): o Web Crawler (Scrapy) procura as informações úteis presentes na página da empresa solicitada;
- Extração dos dados (4): os dados selecionados são capturados;
- Armazenamento dos dados (5): As informações são armazenadas em um banco de dados à medida que são capturadas;
- Obtenção dos dados (6): o Cliente é redirecionado para a página que exibe todas as informações coletadas em uma tabela com três colunas, sendo elas Descrição (um comentário de um consumidor sobre a empresa), Classificação (Elogio ou Reclamação) e Data (quando o comentário foi postado);
- Sumarização dos dados (7): o Cliente precisa clicar no botão MODELAGEM EM TÓPICOS para ativar o algoritmo LDA e ao final de sua execução, o sistema exibe uma página com uma tabela que contém as palavras que mais aparecem nos comentários dos consumidores e a frequência com que aparecem, permitindo ao Cliente verificar se há um número maior de palavras positivas ou negativas, além de descobrir outras informações.
- Resposta (8): o sistema exibe na tela os dados extraídos e a sua sumarização.

## 4 RESULTADOS

A modelagem de tópicos foi realizada para os dados coletados de três empresas. Elas serão identificadas como Empresa A, Empresa B e Empresa C. Na Tabela 1 temos o resumo das informações das três empresas.

| Empresa | Setor | Comentários Coletados |
|---|---|---|
| Empresa A | Bens de Consumo e Serviços de Viagem | 6.000 |
| Empresa B | Bens de Consumo | 6.000 |
| Empresa C | Eletrodomésticos | 6.000 |

**Tabela 1 - Dados do estudo de caso**

O algoritmo LDA foi executado de modo a gerar cinco tópicos de seis palavras para cada uma delas. A Tabela 2 exibe as palavras mais presentes nos tópicos das empresas. Os termos com frequência menor que 2% foram desconsiderados por sua frequência ser irrelevante.

| Empresa A | | | | | |
|---|---|---|---|---|---|
| Tópico | Termos mais frequentes | | | | |
| 0 | preço | loja | produto | boa | compra | facilidade |
| 1 | frete | produto | grátis | | | |
| 2 | bom | compra | gostei | comprar | Empresa A | |
| 3 | entrega | produto | prazo | sempre | excelente | loja |
| 4 | Empresa A | loja | sempre | problema | site | compro |
| **Empresa B** | | | | | |
| Tópico | Termos mais frequentes | | | | |
| 0 | prazo | entrega | produto | antes | | |
| 1 | frete | desconto | livro | compra | loja | opção |
| 2 | compra | site | fácil | parabéns | facilidade | |
| 3 | preço | loja | produto | entrega | Empresa B | excelente |
| 4 | loja | boa | bom | Empresa B | produto | compra |
| **Empresa C** | | | | | |
| Tópico | Termos mais frequentes | | | | |
| 0 | desconto | | | | | |
| 1 | compra | produto | facilidade | site | fácil | comprar |
| 2 | entrega | prazo | compra | produto | frete | gostei |
| 3 | Empresa C | produto | compra | marca | confiança | |
| 4 | Bom | preço | produto | excelente | ótimo | site |

**Tabela 2 - Termos mais frequentes para as empresas estudadas**

Ao analisar alguns dos tópicos das empresas estudadas, podemos juntar algumas palavras e então, extrair ideias que fazem sentido e conseguem resumir qual a opinião dos consumidores sobre estas empresas. Além disso, podemos observar se um tópico possui em sua maioria palavras positivas ou negativas, sendo possível deduzir como é a reputação da empresa. A Tabela 3 mostra as conclusões que podem ser tiradas de combinações das palavras de cada tópico das três empresas.

| Empresa | Tópico | Palavras analisadas | Deduções |
|---|---|---|---|
| A | 0 | *boa, facilidade* | É possível ter uma boa impressão da empresa. |
| | 1 | *compra, frete, grátis* | Para algumas compras, o frete é grátis |
| | 2 | *bom, gostei* | É possível ter uma boa impressão da empresa. |
| | 3 | *entrega, produto, prazo, excelente, loja* | Os produtos são entregues no prazo. A loja é excelente. |
| B | 0 | *prazo, entrega, produto, antes* | Alguns produtos são entregues antes do prazo |
| | 1 | *desconto, compra* | A loja oferece desconto para algumas compras |
| | 2 | *fácil, parabéns, facilidade* | É possível ter uma boa impressão da empresa. |
| | 3 | *Excelente* | É possível ter uma boa impressão da empresa. |
| | 4 | *boa, loja, bom, produto* | É possível ter uma boa impressão da empresa. |
| C | 0 | *desconto, compra, produto* | A loja oferece desconto para algumas compras. |
| | 1 | *Fácil* | É possível ter uma boa impressão da empresa. |
| | 2 | *entrega, prazo, gostei* | Os produtos comprados são entregues no prazo. É possível ter uma boa impressão da empresa. |
| | 3 | *confiança, loja* | A loja é confiável. |
| | 4 | *bom, preço, produto, excelente, ótimo* | Elogios aos preços e produtos da loja. |

**Tabela 3 - Deduções a partir da análise dos tópicos obtidos.**

Com relação ao tópico 4 da Empresa A, não foi possível tirar nenhuma conclusão, uma vez que as palavras em conjunto não fazem sentido. A palavra "problema" poderia ser analisada. No entanto, ela é ambígua. Pode significar que alguns clientes relataram que nunca tiveram problema com a loja, ao passo que outros registraram reclamações, sendo difícil tirar conclusões.

## 5 CONCLUSÕES

Os estudos realizados no presente trabalho mostraram que os consumidores estão cada vez mais engajados nas redes sociais, em busca de informações sobre a reputação de empresas e a qualidade dos bens e serviços por elas oferecidos. Portanto, estes consumidores podem usar os sistemas de reputação de forma passiva, apenas analisando tais informações, e de forma ativa, registrando as suas experiências de consumo e o seu grau de satisfação com as empresas.

Dessa forma, as organizações precisam acompanhar este avanço tecnológico, de modo a encontrar novas formas de se relacionar com seus atuais e futuros clientes. Uma prática que pode ajudar tanto os consumidores quanto as empresas em suas atitudes é a análise automática dos volumosos conjuntos de dados presentes nas plataformas de boca a boca eletrônico. Tendo como base esta linha de raciocínio, foi desenvolvida uma ferramenta capaz de capturar e analisar os dados das empresas cadastradas na plataforma eBit. Esta ferramenta pode ser adaptada para qualquer plataforma de boca a boca eletrônico.

Para criar essa ferramenta, implementou-se uma interface web para que qualquer pessoa fosse capaz de utilizar o rastreador web combinado ao algoritmo LDA, sem precisar entender o processo interno. Com a sumarização dos dados coletados, tornou-se possível conhecer a reputação das empresas e algumas outras características específicas sem ser necessário ler todas as informações a respeito delas. Apenas uma pequena parte dos resultados alcançados nesta pesquisa não correspondeu às expectativas.

## 6  RECOMENDAÇÕES

Como trabalhos futuros, pode-se fazer um estudo mais minucioso do algoritmo LDA, a fim de aumentar a sua precisão e reduzir os resultados que não fazem sentido para o consumidor. Pode-se, também, fazer um estudo mais detalhado para que os resultados da modelagem de tópicos sejam mais legíveis para as pessoas comuns e possam ser exibidos em gráficos. Outra possível abordagem é combinar a modelagem de tópicos com as análises de emoção e sentimentos, com o intuito de obter resultados mais precisos e confiáveis.